\documentclass[prd,aps,tightenlines,a4paper,12pt]{revtex4}

\usepackage{graphicx}
\usepackage{bm}
\usepackage{url}
\usepackage{amsfonts}
\usepackage{amsmath}

\newcommand{\ve}[1]{\mbox{\boldmath$#1$}}

\begin{document}

\title{Gravitational field of one uniformly moving extended body 
and N arbitrarily moving pointlike bodies 
in post-Minkowskian approximation}  

\author{Sven \surname{Zschocke}, Michael H. Soffel}
\affiliation{
Lohrmann Observatory, Dresden Technical University,\\
Helmholtzstrasse 10, \\
 D-01062 Dresden, Germany\\
}

\begin{abstract}

\begin{center}

\medskip


\end{center}

High precision astrometry, space missions and certain tests of General
Relativity, require the knowledge of the metric tensor of the solar system,
or more generally, of a gravitational system of N extended bodies. 
Presently, the metric of arbitrarily shaped, rotating, oscillating and arbitrarily moving
N bodies of finite extension is only known for the case of slowly moving bodies
in the post-Newtonian approximation, 
while the post-Minkowskian metric for arbitrarily
moving celestial objects is known only for pointlike bodies with mass-monopoles
and spin-dipoles.

As one more step towards the aim of a global metric for
a system of N arbitrarily shaped and arbitrarily moving massive bodies in
post-Minkowskian approximation,  
two central issues are on the scope of our investigation: 

(i) We first consider one extended body
with full multipole structure in uniform motion in some suitably chosen
global reference system. For this problem a co-moving inertial system of
coordinates can be introduced where the metric, outside the body, admits an
expansion in terms of Damour-Iyer moments. A Poincar\'e transformation then
yields the corresponding metric tensor in the global system in post-Minkowskian approximation. 

(ii) It will be argued why the global metric, exact to
post-Minkowskian order, can be obtained by means of an instantaneous
Poincar\'e transformation for the case of pointlike mass-monopoles and
spin-dipoles in arbitrary motion.

\end{abstract}


\maketitle

\newpage

\section{Introduction}

Since exact solutions of Einstein's field equations are available only for
highly idealized systems usually one is forced to resort to approximation
schemes. One of the most powerful and most important approximation schemes is
linearized gravity, where the field equations in harmonic coordinates are
simplified to an inhomogeneous wave equation \cite{MTW,Landau_Lifschitz}. As
it has been shown in \cite{Thorne,Blanchet_Damour1,Blanchet_Damour2} in the
post-Newtonian approximation (weak-field slow-motion approximation) 
the metric outside the matter distribution can
be expanded in terms of two families of multipole moments: mass multipole
moments $M_L$ and spin multipole moments $S_L$. Later, in post-Minkowskian approximation  
(weak-field approximation) such a set of 
multipole moments has been introduced by {\it Damour} $\&$ {\it Iyer} \cite{Multipole_Damour_2}.

For many purposes, for instance for high precision astrometry or fundamental
tests of relativity, the knowledge of the global metric of an N-body system
in post-Minkowskian approximation is of fundamental importance. Presently the
post-Minkowskian metric for arbitrarily moving celestial objects is known
only for pointlike bodies with mass-monopoles and spin-dipoles. The metric
of arbitrarily shaped, rotating, oscillating and moving bodies is a highly
sophisticated and complex problem and is only known for the case of slowly
moving bodies in the post-Newtonian approximation \cite{DSX1}. One reason for
this complexity is, that one might want to define the multipole moments of a
single body in its own rest-frame, with origin close to the body's center of
mass; however, if the acceleration of such a 'local' co-moving system is
taken into account corresponding multipole moments have been
defined only to post-Newtonian order \cite{DSX1,DSX2}.

Thus, in order to study the global metric field in terms of locally defined
multipoles of a realistic N-body system such as the solar system, one has to
apply further approximations. Accordingly, this will be the strategy of
this paper: we will first consider an arbitrarily shaped, rotating and
oscillating body first in uniform motion, and then we treat the problem of  
N arbitrarily moving pointlike bodies with mass-monopoles and spin-dipoles.

The article is organized as follows: the metric for an extended body with arbitrary
Damour-Iyer moments, defined in a co-moving system, in uniform motion is
derived in section \ref{Metric_Uniform_Motion} in post-Minkowskian approximation.
In section \ref{Monopole} we consider the post-Minkowskian metric for N arbitrarily moving
pointlike bodies (mass-monopoles and spin-dipoles) and show that our results agree 
with corresponding results from the literature.  
Throughout the article we use fairly standard notation:   

\begin{enumerate}
\item[$\bullet$] $G$ is the gravitational constant and $c$ is the speed of light.  
\item[$\bullet$] Lower case Latin indices $a,b,...$ take values $1,2,3$. 
\item[$\bullet$] Lower case Greek indices $\alpha,\beta,...$ take values $0,1,2,3$. 
\item[$\bullet$] repeated Greek indices mean Einstein summation from $0$ to $3$. 
\item[$\bullet$] $\delta_{ab}=\delta^{ab}=\delta_{a}^{b}={\rm diag}\left(+1,+1,+1\right)$ 
is the three-dimensional Kronecker delta. 
\item[$\bullet$] $\delta_{\alpha\beta}=\delta^{\alpha\beta}=\delta_{\alpha}^{\beta} =  
{\rm diag}\left(+1,+1,+1,+1\right)$ is the four-dimensional Kronecker delta. 
\item[$\bullet$] $\epsilon_{abc}$ is the Levi-Civita symbol. 
\item[$\bullet$] $L$ is a Cartesian multi-index, that means $L=a_1 \dots a_l$. 
\item[$\bullet$] $\gamma=\left(1-v^2/c^2\right)^{-1/2}$ is the Lorentz factor. 
\item[$\bullet$] parentheses surrounding a group of Roman indices mean symmetrization with respect 
to these indices: $\displaystyle A^{(a b)}=\frac{1}{2}\left(A^{a b}+A^{b a}\right)$. 
\item[$\bullet$] $\eta_{\mu\nu} = \eta^{\mu\nu} = {\rm diag}\left(-1,+1,+1,+1\right)$
is the metric tensor of Minkowski space. 
\end{enumerate}

\section{A uniformly moving body: multipole expansion to post-Minkowskian order}\label{Metric_Uniform_Motion}

\subsection{Multipole expansion for a body at rest}

Consider a single massive body in some inertial system of harmonic coordinates
$X^{\mu} = \left(c T \,,\, \ve{X}\right)$. For weak gravitational fields the
metric differs only slightly from flat space metric, that means  
$G_{\mu \nu} = \eta_{\mu \nu} + H_{\mu \nu}$,  
where $\left|H_{\mu \nu}\right| \ll 1$; the metric signature is $\left(-, +, +, +\right)$.  
Weak gravitational fields are governed by the equations of linearized gravity, in harmonic gauge  
given by \cite{MTW} (from now on all relations will be valid to first order in $G$, even if this 
is not indicated explicitly):  
\begin{eqnarray}
\square_X \overline{H}^{\mu\nu} \left(T, \ve{X}\right) &=&
- \frac{16\,\pi\,G}{c^4}\;T^{\mu\nu} \left(T, \ve{X}\right),
\label{Linearized_Gravity}
\end{eqnarray}

\noindent
where $\displaystyle \square_X = \eta^{\mu\nu}\,\frac{\partial^2}{\partial X^{\mu}\,\partial X^{\nu}}$
is the d'Alembert operator, the stress-energy tensor of matter is $T^{\mu\nu}$, and  
$\overline{H}_{\mu \nu}$ is the trace-reversed metric perturbation;   
definitions and relations are given in Appendix \ref{Metric_Notation}.  

{\it Damour} $\&$ {\it Iyer} \cite{Multipole_Damour_2} succeeded to show that outside the
body the metric in (skeletonized) harmonic gauge admits an expansion in terms of two families 
of multipole moments: mass-moments $M_L$ and spin-moments $S_L$. Their canonical  
form of the metric perturbation in the exterior region of the matter field can be written as follows:  
\begin{eqnarray}
H^{\mu \nu}_{\rm can} \left(T,\ve{X}\right) &=&
+\,\frac{2\,G}{c^2}\,\delta_{\mu \nu}\,\sum\limits_{l \ge 0}\,\frac{\left(-1\right)^l}{l!}
{\cal D}_L \left[\frac{M_L \left(T_{\rm ret}\right)}{R} \right]
\nonumber\\
\nonumber\\
&& -\,\frac{8\,G}{c^3}\,\delta_{0\, (\mu}\,\delta_{\nu)\,i}\,\sum\limits_{l \ge 1}\,\frac{\left(-1\right)^l}{l!}\,
{\cal D}_{L-1}\left[\frac{\dot{M}_{i\,L-1}\left(T_{\rm ret}\right)}{R}\right]
\nonumber\\
\nonumber\\
&& -\,\frac{8\,G}{c^3}\,\delta_{0\, (\mu}\,\delta_{\nu)\,k}\,
\sum\limits_{l \ge 1}\,\frac{\left(-1\right)^l\,l}{\left(l+1\right)!}\,
\epsilon_{i\,j\,k}\,{\cal D}_{i\,L-1}\left[\frac{S_{j\,L-1}\left(T_{\rm ret}\right)}{R}\right] 
\nonumber\\
\nonumber\\
&& +\,\frac{4\,G}{c^4}\,\delta_{\mu i}\,\delta_{\nu j}\,\sum\limits_{l \ge 2}\,\frac{ \left(-1\right)^l}{l!}
{\cal D}_{L-2}\,\left[\frac{\ddot{M}_{i j\,L-2}\left(T_{\rm ret}\right)}{R}\right]
\nonumber\\
\nonumber\\
&& + \,\frac{8\,G}{c^4}\,\delta_{\mu i}\,\delta_{\nu j}\,\sum\limits_{l \ge 2}\,
\frac{\left(-1\right)^l\,l}{\left(l+1\right)!}\,{\cal D}_{k\,L-2}\,\left[\frac{\epsilon_{k m\,(\,i}\,\dot{S}_{j\,)\,m\,L-2}
\left(T_{\rm ret}\right)}{R}\right]. 
\label{Local_Metric}
\end{eqnarray}

\noindent
In Eq.~(\ref{Local_Metric}), an  
overdot denotes the derivative with respect to $T_{\rm ret}$;   
e.g. $\displaystyle \dot{F}\left(T_{\rm ret}\right)=\frac{d F\left(T_{\rm ret}\right)}{d T_{\rm ret}}$
for any function $F$, and  
all multipole moments are taken at the retarded instance of time, 
\begin{eqnarray}
T_{\rm ret} \left(T,\ve{X}\right) &=& T - \frac{R}{c}\,,
\label{retarded_time}
\end{eqnarray}

\noindent 
with $R=\left|\ve{X}\right|$. The multipole moments, $M_L$ and
$S_L$, are Cartesian symmetric and trace-free (STF) tensors;
$\displaystyle {\cal D}_L = \partial^l/(\partial X^{a_1}\,\partial X^{a_2}\,...\,\partial X^{a_l})$.  
Explicit expressions for the multipole moments, $M_L$ and  $S_L$,
in post-Minkowskian approximation are given by Eqs.~(5.33) - (5.35) in
\cite{Multipole_Damour_2}.

\subsection{Multipole expansion for a uniformly moving body}

\noindent
Considering a single body in uniform motion we will now attach our inertial coordinates $X^\mu = \left(cT\,,\,\ve{X}\right)$ to
the body, by choosing its origin near the body's center of mass. The spatial coordinate $\ve{X}^{\rm CoM} = 0$ of
center of mass can be defined by the vanishing of the corresponding Damour-Iyer mass-dipole moment $\ve{M}_a=0$,  
but we consider the more general case with $\ve{M}_a \neq 0$ and $\ve{X}^{\rm CoM} \neq 0$. This coordinate
system will be called co-moving in the following (or 'local' in case that the body's velocity is time dependent).

We now consider another inertial (global) system of coordinates $x^{\mu} = \left(ct\,,\,\ve{x}\right)$ in which our 
body moves with constant velocity $\ve{v}$. The transformation from local coordinates $X^{\mu} = \left(cT,\ve{X}\right)$ 
to global coordinates $x^{\mu} = \left(ct,\ve{x}\right)$ for a massive body in uniform motion is given by a Poincar\'e 
transformation,
\begin{eqnarray}
x^{\mu} \left(X^{\alpha}\right) &=& b^{\mu} + \Lambda^{\mu}_{\alpha}\,X^{\alpha}\,,
\label{Poincare_transformation}
\end{eqnarray}

\noindent
with $\Lambda^0_0 = \gamma$,
$\Lambda^i_0 = \Lambda_i^0 = \gamma\,\frac{\displaystyle v_i}{\displaystyle c}$, $\Lambda_i^j =
\delta_{ij} + \left(\gamma-1\right)\frac{\displaystyle v_i\,v_j}{\displaystyle v^2}$, and 
$b^{\mu} = \left(b^0, \ve{b}\right)$ is a constant four-vector, where $\ve{b}$ points 
from the origin ${\cal O}$ of global frame to the origin of the co-moving frame at time $T=0$. 
Transforming the events $\left(T,\ve{0}\right)$ into the global reference system $\left(t,\ve{x}\right)$ yields 
\begin{eqnarray}
\ve{x}_A \left(t\right) &=& \ve{x}_A \left(t_0\right) + \ve{v} \left(t - t_0\right), \quad 
x_A^i \left(t_0\right) = b^i\,, \quad c\,t_0 = b^0\,, 
\label{retarded_time_E}
\end{eqnarray}

\noindent 
where $\ve{x}_A\left(t\right)$ points from the origin of the global system to the origin at the co-moving frame at 
any time $t$, and the initial is $t_0$. The distance $R$ which appears in the co-moving 
metric (\ref{Local_Metric}), can be written in Lorentz invariant form $\rho$ as
(cf. Eq.~(4.42) in \cite{DSX1}, Eq.~(10) in \cite{KopeikinSchaefer1999}, Eq.~(B.4) in \cite{Klioner_A&A})
\begin{eqnarray}
\rho &=& \frac{\left|\eta_{\mu\nu}\, u^{\mu}\,\left(x^{\nu} - x_A^{\nu}\left(t_{\rm ret}\right) \right)\right|}{c}\,,
\label{invariant_distance_1}
\end{eqnarray}

\noindent
where $u^{\mu} = \gamma \left(c,\ve{v}\right)$ are the contravariant components of four-velocity 
of ${\cal O}$,
and the retarded time is defined by Eq.~(\ref{retarded_time_D}) below. 
The Lorentz invariant distance (\ref{invariant_distance_1}) can also be written as:
\begin{eqnarray}
\rho &=& \gamma\,
\left( r\left(t_{\rm ret}\right) - \frac{\ve{v} \cdot \ve{r} \left(t_{\rm ret}\right)}{c}\right)
= \sqrt{r^2 \left(t\right) + \gamma^2 \, \frac{\left(\ve{v} \cdot \ve{r} \left(t\right)\right)^2}{c^2}}\,,
\label{invariant_distance_2}
\end{eqnarray}

\noindent
where $\ve{r} \left(t_{\rm ret}\right) = \ve{x} - \ve{x}_A \left(t_{\rm ret}\right)$,
$\ve{r} \left(t\right) = \ve{x} - \ve{x}_A \left(t\right)$ and 
$\ve{x}_A \left(t\right)$ is given by Eq.~(\ref{retarded_time_E}); the absolute values are 
$r \left(t_{\rm ret}\right) = \left|\ve{r} \left(t_{\rm ret}\right)\right|$
and $r \left(t\right) = \left|\ve{r} \left(t\right)\right|$.

The latter form in (\ref{invariant_distance_2}) is sometimes preferable and can be obtained by means of
the relation $X^i = \Lambda^i_a\,r^a \left(t\right)$; for a very similar consideration see \cite{Klioner_A&A}.
The retarded time in global coordinates reads for arbitrary wordlines  
\begin{eqnarray}
t_{\rm ret}\left(t,\ve{x}\right) &=& 
t - \frac{\left| \ve{x} - \ve{x}_A \left(t_{\rm ret} \right)\right|}{c}\,.
\label{retarded_time_D}
\end{eqnarray}

\noindent 
Let us consider a series expansion of (\ref{retarded_time_D}), which yields:
$\displaystyle t_{\rm ret} = t - \frac{r\left(t\right)}{c}
- \frac{\ve{v}\left(t\right)\cdot\ve{r}\left(t\right)}{c^2} + {\cal O} \left(c^{-3}\right)$,
where $\ve{r} \left(t\right) = \ve{x} - \ve{x}_A \left(t\right)$,
and $\ve{x}_A \left(t\right)$ is arbitrarily, hence $\ve{v}\left(t\right)= \dot{\ve{x}}_A \left(t\right)$ 
is time-dependent. In general, Eq.~(\ref{retarded_time_D}) is an implicit relation  
which cannot be resolved analytically for arbitrary worldlines $\ve{x}_A\left(t\right)$ of a massive body.  
However, for the case of a body in uniform motion one can obtain an exact analytical solution:  
\begin{eqnarray}
t_{\rm ret} \left(t,\ve{x}\right) &=& t - \gamma^2\;
\frac{\ve{r}\left(t\right)\cdot\ve{v}+\sqrt{c^2\,r^2\left(t\right)-\left(\ve{r}\left(t\right)\times\ve{v}\right)^2}}{c^2}\,.
\label{retarded_time_G}
\end{eqnarray}

\noindent 
Here, $\ve{r} \left(t\right) = \ve{x} - \ve{x}_A \left(t\right)$,
and $\ve{x}_A \left(t\right)$ is given by Eq.~(\ref{retarded_time_E}).
Let us compare (\ref{retarded_time_G}) with the post-Newtonian approximation.
A series expansion of (\ref{retarded_time_G}) yields the following expression for the 
retarded time:  
$\displaystyle t_{\rm ret} = t - \frac{r\left(t\right)}{c} 
- \frac{\ve{v}\cdot\ve{r}\left(t\right)}{c^2} + {\cal O} \left(c^{-3}\right)$. This expression  
agrees with the series expansion given above (for $\ve{v}={\rm const}$)  
which has been obtained directly 
from the definition (\ref{retarded_time_D}).

Now we consider a relation among the retarded time $T_{\rm ret}$ in the
co-moving system of the body and the retarded time $t_{\rm ret}$ in the
global system. The retarded time in the co-moving and global system are defined
by Eqs.~(\ref{retarded_time}) and (\ref{retarded_time_D}), respectively. In
order to find a relation between $T_{\rm ret}$ and $t_{\rm ret}$, we note
that the global coordinates of event $\left(t_{\rm ret},\ve{x}_A \left(t_{\rm ret}\right)\right)$ correspond 
to the coordinates $\left(T_{\rm ret},\ve{0}\right)$ of the same event in the co-moving frame.
The Poincar\'e transformation of the coordinates of this event yields
\begin{eqnarray}
T_{\rm ret} &=& \gamma^{-1}\left(t_{\rm ret} - t_0\right)\,.
\label{appendix_RT_25}
\end{eqnarray}

\noindent 
Relation (\ref{appendix_RT_25}) can also be obtained directly from
the definitions of $T_{\rm ret}$ and $t_{\rm ret}$.

To get the metric in the global system we will transform the spatial
derivatives with respect to the co-moving coordinates to derivatives with
respect to global coordinates. One obtains
\begin{eqnarray}
{\cal D}_L \left[\frac{F \left(T_{\rm ret}\right)}{R} \right] &=&
\Lambda^{\mu_1}_{a_1}\,...\,\Lambda^{\mu_l}_{a_l}\,
\partial_{\mu_1\,...\,\mu_l}\left[\frac{F\left(\gamma^{-1} \left(t_{\rm ret}-t_0\right)\right)}{\rho\left(t_{\rm ret}\right)}\right]\,,
\label{LT_differential_10}
\end{eqnarray}

\noindent 
where $F$ stands for any of the mass or spin multipoles in co-moving coordinates, and 
$\partial_{\mu} = \partial/\partial x^{\mu}$. 
By means of the invariant 
form of the distance (\ref{invariant_distance_1}) and with the aid of the derivative operation (\ref{LT_differential_10}), 
we are in the position to obtain the global metric in terms of local multipoles for a massive body in uniform motion. 
Using $g^{\mu \nu}=\eta^{\mu \nu} - h^{\mu \nu} + {\cal O} \left(G^2\right)$,
$G^{\alpha \beta}=\eta^{\alpha \beta} - H^{\alpha \beta} + {\cal O} \left(G^2\right)$ and relation
$\Lambda^{\mu}_{\alpha}\,\Lambda^{\nu}_{\beta}\,\eta^{\alpha \beta}=\eta_{\mu \nu}$,
we obtain from (\ref{Poincare_transformation}) the transformation law of metric perturbation:
\begin{eqnarray}
h_{\rm can}^{\mu \nu}\left(t,\ve{x}\right) &=& \Lambda^{\mu}_{\alpha}\,\Lambda^{\nu}_{\beta}\,
H_{\rm can}^{\alpha\beta}\left(T,\ve{X}\right).
\label{coordinate_transformation_8}
\end{eqnarray}

\noindent
Applying the general transformation law (\ref{coordinate_transformation_8}) to the local metric
(\ref{Local_Metric}), using the invariant form of the distance (\ref{invariant_distance_1}),
the derivative operation (\ref{LT_differential_10}), we obtain for the metric in global
coordinates $\left(t,\ve{x}\right)$ the following expression:
\begin{eqnarray}
h^{\mu \nu}_{\rm can} \left(t,\ve{x}\right) &=& \frac{2\,G}{c^2}\,
\Lambda^{\mu}_{\alpha}\,\Lambda^{\nu}_{\beta}\, \delta_{\alpha\beta}\,
\sum\limits_{l \ge 0}\,\frac{\left(-1\right)^l}{l!}\,
\Lambda^{\mu_1}_{a_1}\,...\,\Lambda^{\mu_l}_{a_l}\,
\partial_{\mu_1\,...\,\mu_l}
\left[\frac{M_{a_1\,...\,a_l}\left(T_{\rm ret}\right)}{\rho}\right]
\nonumber\\
\nonumber\\
&& \hspace{-2.0cm} -\,\frac{8\,G}{c^3}\,\Lambda^{(\mu}_0\,\Lambda^{\nu)}_i\,
\sum\limits_{l \ge 1}\,\frac{\left(-1\right)^l}{l!}\,
\Lambda^{\mu_1}_{a_1}\,...\,\Lambda^{\mu_{l-1}}_{a_{l-1}}\,
\partial_{\mu_1\,...\,\mu_{l-1}}
\left[\frac{\dot{M}_{i a_{1}\,...\,a_{l-1}}\left(T_{\rm ret}\right)}{\rho}\right]
\nonumber\\
\nonumber\\
&& \hspace{-2.0cm} -\,\frac{8\,G}{c^3}\,\Lambda^{(\mu}_0\,\Lambda^{\nu)}_k\,
\sum\limits_{l \ge 1}\,\frac{\left(-1\right)^l\,l}{\left(l+1\right)!}\,\epsilon_{i\,j\,k}\,
\Lambda^{\lambda}_{i}\,\Lambda^{\mu_1}_{a_1}\,...\,\Lambda^{\mu_{l-1}}_{a_{l-1}}\,
\partial_{\lambda \mu_1\,...\,\mu_{l-1}}
\left[\frac{S_{j a_1\,...\,a_{l-1}} \left(T_{\rm ret}\right)}{\rho}\right]
\nonumber\\
\nonumber\\
&& \hspace{-2.0cm} +\,\frac{4\,G}{c^4}\,\Lambda^{\mu}_i\,\Lambda^{\nu}_j\,
\sum\limits_{l \ge 2}\,\frac{ \left(-1\right)^l}{l!}\,
\Lambda^{\mu_1}_{a_1}\,...\,\Lambda^{\mu_{l-2}}_{a_{l-2}}\,
\partial_{\mu_1\,...\,\mu_{l-2}}
\left[\frac{\ddot{M}_{i j a_1\,...\,a_{l-2}} \left(T_{\rm ret}\right)}{\rho}\right]
\nonumber\\
\nonumber\\
&& \hspace{-2.0cm} +\,\frac{8\,G}{c^4}\,\Lambda^{\mu}_i\,\Lambda^{\nu}_j
\sum\limits_{l \ge 2}\,\frac{ \left(-1\right)^l\,l}{\left(l+1\right)!}\, 
\Lambda^{\lambda}_{k}\,\Lambda^{\mu_1}_{a_1}\,...\,\Lambda^{\mu_{l-2}}_{a_{l-2}}\,
\partial_{\lambda \mu_1\,...\,\mu_{l-2}}
\left[\frac{\epsilon_{k m\,(\,i}\,\dot{S}_{j\,)\,m\,a_1\,...\,a_{l-2}}\left(T_{\rm ret}\right)}{\rho}\right],
\nonumber\\
\label{Global_Metric_A}
\end{eqnarray}

\noindent 
where $T_{\rm ret}$ can be expressed in terms of global coordinates by means of (\ref{appendix_RT_25}),   
and an overdot denotes the derivative with respect to $T_{\rm ret}$. 
The multipoles in (\ref{Global_Metric_A}) are the local multipoles defined in the co-moving frame of the body under 
consideration, and they are functions of the retarded time $T_{\rm ret}$.  
Expression (\ref{Global_Metric_A}) describes the metric of an
arbitrarily shaped and arbitrarily oscillating and rotating single massive body in uniform motion.

\subsection{Monopole in uniform motion}\label{monopole_uniform}

Let us consider the simplest case of an extended body with monopole structure.
According to Eq.~(\ref{Global_Metric_A}), the metric perturbation of a
uniformely moving mass-monopole in global coordinates $x^{\mu} = \left(c t,
\ve{x}\right)$ is given by ($l=0$ in Eq.~(\ref{Global_Metric_A})):
\begin{eqnarray}
h^{\mu \nu}_{\left(M\right)} \left(t,\ve{x}\right) &=& \frac{2\,G\,M}{c^2}\,
\frac{\Lambda^{\mu}_{\alpha}\;\Lambda^{\nu}_{\beta}\;\delta_{\alpha\beta}}{\rho}\,,
\label{Global_Metric_Pointmass_0}
\end{eqnarray}

\noindent
where $M$ is the invariant rest mass of the body. For the invariant distance $\rho$ we insert 
expression (\ref{invariant_distance_2}), then we use the relation
$c^2\,\Lambda^{\mu}_{\alpha}\;\Lambda^{\nu}_{\beta}\,\delta_{\alpha \beta} = 2\,u^{\mu}\,u^{\nu} + c^2\,\eta^{\mu \nu}$,
and obtain 
\begin{eqnarray}
h^{\mu \nu}_{\left(M\right)} \left(t,\ve{x}\right) &=&
\frac{4\,G\,M}{c^2}\,\frac{1}{\gamma \left(r\left(t_{\rm
ret}\right)-\frac{\displaystyle \ve{v}\cdot\ve{r}\left(t_{\rm
ret}\right)}{\displaystyle c}\right)} \left(\frac{u^{\mu}}{c}
\frac{u^{\nu}}{c} + \frac{1}{2}\,\eta^{\mu \nu}\right). \label{evaluation_45}
\end{eqnarray}

\noindent
This expression coincides with the metric of a pointlike body of mass $M$, cf. Eq.~(\ref{evaluation_45})
with Eq.~(11) in \cite{KopeikinMashhoon2002} for the case of uniform motion, i.e. $\ve{v} = {\rm const}$.

\subsection{Spin-dipole in uniform motion}\label{spin_uniform}

Next we consider an extended massive body with mass monopole $M$ and spin dipole $S_i$.
According to Eq.~(\ref{Global_Metric_A}), the metric perturbation of a
uniformely moving mass-monopole in global coordinates $x^{\mu} = \left(c t,
\ve{x}\right)$ consists of two contributions ($l=1$ in
Eq.~(\ref{Global_Metric_A})):
\begin{eqnarray}
h^{\mu \nu} \left(t,\ve{x}\right) &=&
h^{\mu \nu}_{\left(M\right)} \left(t,\ve{x}\right) + h^{\mu \nu}_{\left(S\right)} \left(t,\ve{x}\right),
\label{Spin_5}
\end{eqnarray}

\noindent
where for simplicity we assume in this section that the co-moving system is located at the
center of mass of this body, so that $M_a = 0$.
The monopole part $h^{\mu \nu}_{\left(M\right)}$ is given by Eq.~(\ref{evaluation_45}), and the spin part
$h^{\mu \nu}_{\left(S\right)}$, according to (\ref{Global_Metric_A}), is given by
\begin{eqnarray}
h^{\mu \nu}_{\left(S\right)} \left(t,\ve{x}\right) &=&
\frac{4\,G}{c^3}\;\Lambda^{(\mu}_0\;\Lambda^{\nu)}_k\,\epsilon_{i\,j\,k}\;\Lambda^{\lambda}_i\;
\partial_{\lambda} \,\frac{S_j}{\rho}\,,
\label{Spin_10}
\end{eqnarray}

\noindent
where $S_j$ is the spin in the local frame $\left(cT,\ve{X}\right)$ of the body.

The massive bodies of an N-body system exert a torque on each other leading
to a time dependent spin of a body A in the local A-system. Here, we follow
the arguments of \cite{Martinez_A,Martinez_B} and will assume that such a
local time-dependence is only caused by gravitational interactions and,
therefore, are proportional to ${\cal O} \left(G\right)$. Accordingly, the
spin of each individual body in its own co-moving system is here assumed to
be time independent. The metric of an arbitrarily moving pointlike body with
monopole structure and a time-independent spin has been given by Eq.~(16)
in \cite{KopeikinMashhoon2002}. Here, we will compare our result
with the results in \cite{KopeikinMashhoon2002} in case of a body in uniform motion.

Because the spin is time-independent in the local frame, 
the derivative $\partial_{\lambda}$ in (\ref{Spin_10})
does not act on the spin vector, and we obtain
\begin{eqnarray}
\Lambda^{\alpha}_i\;\partial_{\alpha} \,\frac{S_j}{\rho} &=& - S_j\,\frac{r^i \left(t\right)+\left(\gamma-1\right)\,
\frac{\displaystyle v^i}{\displaystyle v^2}\,\left(\ve{v}\cdot\ve{r}\left(t\right)\right)}{\rho^3}\;.
\label{Spin_11}
\end{eqnarray}

\noindent
In order to obtain (\ref{Spin_11}), we have used the second expression in (\ref{invariant_distance_2}), the explicit form
for the Lorentz matrices, and $\frac{\displaystyle \partial}{\displaystyle \partial t} \ve{r} \left(t\right) = - \ve{v}$.
By inserting (\ref{Spin_11}) into (\ref{Spin_10}) we obtain
\begin{eqnarray}
h^{\mu \nu}_{\left(S\right)} \left(t,\ve{x}\right) &=&
- \frac{4\,G}{c^3}\;\Lambda^{(\mu}_0\;\Lambda^{\nu)}_k\,\epsilon_{i\,j\,k}\;\,S_{j}\;
\frac{r^i \left(t\right)+\left(\gamma-1\right)\,
\frac{\displaystyle v^i}{\displaystyle v^2}\,\left(\ve{v}\cdot\ve{r}\left(t\right)\right)}{\rho^3}\,.
\label{Spin_12}
\end{eqnarray}

\noindent
Furthermore, we note the relation
\begin{eqnarray}
r^i \left(t\right) &=& r^i \left(t_{\rm ret}\right) - r \left(t_{\rm ret}\right)\,\frac{v^i}{c}\,,
\label{Spin_13}
\end{eqnarray}

\noindent
which follows from $r^i \left(t\right) = x^i - x_A^i \left(t\right)$,
$r^i \left(t_{\rm ret}\right) = x^i - x_A^i \left(t_{\rm ret}\right)$, that means
$r^i \left(t\right) = r^i \left(t_{\rm ret}\right) + v^i \left(t_{\rm ret} - t\right)$, and then using
relation (\ref{retarded_time_D}). Thus, by means of (\ref{Spin_13}), we can rewrite (\ref{Spin_12}) as follows:
\begin{eqnarray}
h^{\mu \nu}_{\left(S\right)} \left(t,\ve{x}\right) &=&
- \frac{4\,G}{c^3}\;\Lambda^{(\mu}_0\;\Lambda^{\nu)}_k\,\epsilon_{i\,j\,k}\;\,S_{j}\;
\frac{r^i \left(t_{\rm ret}\right)+\left(\gamma-1\right)\,
\frac{\displaystyle v^i}{\displaystyle v^2}\,\left(\ve{v}\cdot\ve{r}\left(t_{\rm ret}\right)\right)
- \gamma\,r\left(t_{\rm ret}\right)\frac{\displaystyle v^i}{\displaystyle c}}{\rho^3}\,.
\label{Spin_14}
\end{eqnarray}

\noindent
Let us briefly note, that from (\ref{Spin_14}) one easily finds that $h^{00}_{\left(S\right)}={\cal O} \left(c^{-4}\right)$,
$h^{ij}_{\left(S\right)}={\cal O} \left(c^{-4}\right)$, while
\begin{eqnarray}
h^{0i}_{\left(S\right)} \left(t,\ve{x}\right) &=& \frac{2\,G}{c^3}\,\epsilon_{ijk}\,S_j\,
\frac{r_k\left(t\right)}{r^3\left(t\right)} + {\cal O} \left(c^{-5}\right)
\label{Spin_15}
\end{eqnarray}

\noindent 
gives rise to the famous Lense-Thirring effect.  
Now we will show an agreement of (\ref{Spin_14}) with
Eq.~(16) in \cite{KopeikinMashhoon2002}, where  some anti-symmetric spin
tensor in global coordinates $S^{\alpha \beta}_{\rm global}$ has been
employed. To this end we consider each component of the metric tensor  
(\ref{Spin_14}) separately.

Accordingly, the strategy for the comparison is, first to perform a Lorentz
transformation of the spin-part of the metric from co-moving to global
frame and second to rewrite the results in terms of the global spin
tensor $S^{\mu\nu}_{\rm global}$, Eqs.~(\ref{appendix_spin_relations_5}) and
(\ref{appendix_spin_relations_6}).

\subsubsection{Calculation of  $h^{00}_{\left(S\right)}$}

For the components $\mu=\nu=0$ we obtain from (\ref{Spin_14}) the following expression:
\begin{eqnarray}
h^{0 0}_{\left(S\right)} \left(t,\ve{x}\right) &=& \frac{4\,G}{c^3}\,\gamma^2\,
\left(\frac{\ve{v}}{c} \times \ve{S}\right)^i\,\frac{r_i \left(t_{\rm ret}\right)}{\rho^3}\,. 
\label{Spin_00_5}
\end{eqnarray}

\noindent
Now we use the following relation between
the spin vector $\ve{S}$ in the co-moving system and the anti-symmetric spin tensor $S^{\alpha\beta}_{\rm global}$
in the global system, which is shown in Appendix \ref{Spin_Tensor_Relations}:
\begin{eqnarray}
\gamma\left(\frac{\ve{v}}{c} \times \ve{S}\right)^i &=& S^{i 0}_{\rm global}\,.
\label{Spin_00_10}
\end{eqnarray}

\noindent
Inserting (\ref{Spin_00_10}) into (\ref{Spin_00_5}) yields
\begin{eqnarray}
h^{0 0}_{\left(S\right)} \left(t,\ve{x}\right) &=&
\frac{4\,G}{c^3}\,
\frac{r_{\alpha} \left(t_{\rm ret}\right)\;S^{\alpha 0}_{\rm global}\,u^0}{\rho^3}\,,
\label{Spin_00_15}
\end{eqnarray}

\noindent where the four-vector $r^{\alpha} = \left(r,\ve{r}\right)$ has been
introduced. In (\ref{Spin_00_15}) we have formally extended the summation
$r_i\,S^{i 0}_{\rm global} = r_{\alpha}\,S^{\alpha 0}_{\rm global}$, because
$S^{0 0}_{\rm global} = 0$ due to the anti-symmetry of the spin tensor; note
also $\gamma =  u^0/c$ and   
$S^{\alpha 0}_{\rm global}\,u^0 = S^{\alpha ( 0}_{\rm global}\,u^{0)}$.

\subsubsection{Calculation of  $h^{0\,a}_{\left(S\right)}$}

Now let us consider the component $\mu=a$ and $\nu=0$ in (\ref{Spin_14}), which we separate into two terms
as follows,
\begin{eqnarray}
h^{a 0}_{\left(S\right)} \left(t,\ve{x}\right) &=& h^{a 0}_1
\left(t,\ve{x}\right) + h^{a 0}_2 \left(t,\ve{x}\right), \label{Spin_a0_5}
\\
\nonumber\\
h^{a 0}_1 \left(t,\ve{x}\right) &=& -
\frac{2\,G}{c^3}\;\Lambda^a_0\;\Lambda^{0}_k\,\epsilon_{i\,j\,k}\;\,S_{j}\;
\frac{r^i \left(t_{\rm ret}\right)+\left(\gamma-1\right)\,
\frac{\displaystyle v^i}{\displaystyle
v^2}\,\left(\ve{v}\cdot\ve{r}\left(t_{\rm ret}\right)\right) -
\gamma\,r\left(t_{\rm ret}\right)\frac{\displaystyle v^i}{\displaystyle
c}}{\rho^3}\,, \label{Spin_a0_10}
\\
\nonumber\\
h^{a 0}_2 \left(t,\ve{x}\right) &=& -
\frac{2\,G}{c^3}\;\Lambda^0_0\;\Lambda^{a}_k\,\epsilon_{i\,j\,k}\;\,S_{j}\;
\frac{r^i \left(t_{\rm ret}\right)+\left(\gamma-1\right)\,
\frac{\displaystyle v^i}{\displaystyle
v^2}\,\left(\ve{v}\cdot\ve{r}\left(t_{\rm ret}\right)\right) -
\gamma\,r\left(t_{\rm ret}\right)\frac{\displaystyle v^i}{\displaystyle
c}}{\rho^3}\,. \label{Spin_a0_15}
\end{eqnarray}

\noindent
For the expression (\ref{Spin_a0_10}) we obtain
\begin{eqnarray}
h^{a 0}_1 \left(t,\ve{x}\right) &=& \frac{2\,G}{c^4}\;
\frac{r_{\alpha}\left(t_{\rm ret}\right)\,S^{\alpha 0}_{\rm
global}\,u^{a}}{\rho^3}\;, \label{Spin_a0_20}
\end{eqnarray}

\noindent
where we used (\ref{Spin_00_10}) 
and again extended the summation $r_i\,S^{i 0}_{\rm global} = r_{\alpha}\,S^{\alpha 0}_{\rm global}$;
note also $u^a = \gamma\,v^a$. For the term (\ref{Spin_a0_15}) we obtain
\begin{eqnarray}
h^{a 0}_2 \left(t,\ve{x}\right) &=& \frac{2\,G}{c^4}\;
\frac{r_{\alpha}\left(t_{\rm ret}\right)\,S^{\alpha a}_{\rm
global}\,u^{0}}{\rho^3}\;. \label{Spin_a0_25}
\end{eqnarray}

\noindent 
The proof of relation (\ref{Spin_a0_25}) is a bit involved; it can
be found in Appendix \ref{Proof_Relation_Spin_a0_25}. According to
(\ref{Spin_a0_5}) we add both terms (\ref{Spin_a0_20}) and (\ref{Spin_a0_25})
together, and obtain by means of symmetrization notation:
\begin{eqnarray}
h^{a 0}_{\left(S\right)} \left(t,\ve{x}\right) &=& \frac{4\,G}{c^4}\;
\frac{r_{\alpha}\left(t_{\rm ret}\right)\,S^{\alpha ( a}_{\rm global}\,u^{0 )}}{\rho^3}\;.
\label{Spin_a0_30}
\end{eqnarray}

\noindent
We remark that $h^{a 0}_{\left(S\right)} = h^{0 a}_{\left(S\right)}$ as it follows from (\ref{Spin_14}).

\subsubsection{Calculation of  $h^{ab}_{\left(S\right)}$}

According to (\ref{Spin_14}), we obtain the following components for the spin part of metric tensor,
\begin{eqnarray}
h^{a b}_{\left(S\right)} \left(t,\ve{x}\right) &=&
- \frac{4\,G}{c^3}\;\Lambda^{(a}_0\;\Lambda^{b)}_k\,\epsilon_{i\,j\,k}\;\,S_{j}\;
\frac{r^i \left(t_{\rm ret}\right)+\left(\gamma-1\right)\,
\frac{\displaystyle v^i}{\displaystyle v^2}\,\left(\ve{v}\cdot\ve{r}\left(t_{\rm ret}\right)\right)
- \gamma\,r\left(t_{\rm ret}\right)\frac{\displaystyle v^i}{\displaystyle c}}{\rho^3}\,.
\label{Spin_ab_5}
\end{eqnarray}

\noindent 
If we compare expression (\ref{Spin_ab_5}) with expression
(\ref{Spin_a0_15}), we recognize that:
\begin{eqnarray}
h^{a b}_{\left(S\right)} \left(t,\ve{x}\right) &=& h^{a 0}_2
\left(t,\ve{x}\right)\,\frac{v^b}{c} + h^{b 0}_2
\left(t,\ve{x}\right)\,\frac{v^a}{c}\,. \label{Spin_ab_10}
\end{eqnarray}

\noindent
In view of relation (\ref{Spin_ab_10}) and by means of (\ref{Spin_a0_25}), we immediately conclude
\begin{eqnarray}
h^{a b}_{\left(S\right)} \left(t,\ve{x}\right) &=& \frac{4\,G}{c^4}\;
\frac{r_{\alpha}\left(t_{\rm ret}\right)\,S^{\alpha ( a}_{\rm global}\,u^{b )}}{\rho^3}\,.
\label{Spin_ab_15}
\end{eqnarray}

\subsubsection{Collection of terms}

Now we collect the results (\ref{Spin_00_15}), (\ref{Spin_a0_30}) and (\ref{Spin_ab_15}) together and obtain finally
\begin{eqnarray}
h^{\mu \nu}_{\left(S\right)} \left(t,\ve{x}\right) &=& \frac{4\,G}{c^4}\;
\frac{r_{\alpha}\left(t_{\rm ret}\right)\,S^{\alpha ( \mu}_{\rm global}\,u^{\nu )}}{\gamma^3
\left(r\left(t_{\rm ret}\right)-\frac{\displaystyle \ve{v}\cdot\ve{r}\left(t_{\rm ret}\right)}{\displaystyle c}\right)^3}\,,
\label{Spin_mu_nu}
\end{eqnarray}

\noindent 
where we have used for the distance $\rho$ the form given by
relation (\ref{invariant_distance_2}). The metric (\ref{Spin_mu_nu}) for the
spin part coincides with the metric given by Eq.~(16) in
\cite{KopeikinMashhoon2002} for the case of uniform motion, besides an
additional factor $\gamma^{-1}$ which is missing in Eq.~(16) of
\cite{KopeikinMashhoon2002}, as it has been noted already in
\cite{Bruegmann_Thesis}. We note, that the use of a {\it spin tensor} or {\it
spin vector} is more or less a matter of taste and allows for a more compact
notation, but from the physical point of view it is not important at all.
However, it is important that the metric (\ref{Spin_mu_nu}) is given in terms
of {\it global} spin parameters, while our metric (\ref{Spin_10}) for the
spin is given in terms of {\it local} spin parameters. Here, we have shown
that both expressions are equivalent.

\section{Arbitrarily moving pointlike bodies to post-Minkowskian order}\label{Monopole}

\subsection{Instantaneous Poincar\'e transformation and classical electrodynamics}\label{classical_electrodynamics}

Let us consider the equations of classical electrodynamics in the Lorentz-gauge \cite{Jackson}, 
\begin{eqnarray}
\square_x A^{\mu} \left(t,\ve{x}\right) &=& - \mu_0\,j^{\mu}\left(t,\ve{x}\right),
\label{ED_1}
\end{eqnarray}

\noindent
where $\displaystyle \square_x = \eta^{\mu\nu}\,\frac{\partial^2}{\partial x^{\mu}\,\partial x^{\nu}}$
is the d'Alembert operator, $A^{\mu} = \left(\varphi/c,\ve{A}\right)$ is the  
four-potential with scalar-potential $\varphi$ and vector-potential $\ve{A}$, and
$j^{\mu} = \left(c\,\rho\,,\,\ve{j}\right)$
is the four-current with electric charge density $\rho$ and
electric current density $\ve{j}$; the vacuum permeability $\mu_0$ and vacuum permittivity $\epsilon_0$ 
are related via $c^{-2}= \epsilon_0\,\mu_0$.

The equations of linearized gravity (\ref{Linearized_Gravity}) and the equations of  
classical electrodynamics (\ref{ED_1}) have the same mathematical structure.   
Thus we can use some
arguments of classical electrodynamics for our purposes. Especially, we will show that the
problem of an arbitrarily moving pointlike body in linearized gravity 
is similar to the problem of an arbitrarily 
moving pointlike charge $Q$ in electromagnetism.

Let us consider a pointlike charge Q which in the global inertial system 
$x^{\mu} = \left(ct,\ve{x}\right)$ 
is moving along an arbitrary timelike worldline parametrized by $x_Q^{\mu} \left(T\right)$.  
At each instant of time we   
introduce an inertial system $X^{\mu} = \left(cT,\ve{X}\right)$
along the worldline $x_Q^{\mu}\left(T\right)$ which is comoving with the pointlike charge with 
the instantaneous velocity of the charge. 
The transformation from the
global inertial coordinate system $x^{\mu} = \left(ct,\ve{x}\right)$ to the
inertial system $X^{\mu} = \left(cT,\ve{X}\right)$ which is comoving with the charge
is then given by an instantaneous Poincar\'e
transformation, e.g. \cite{Kopeikin_Efroimsky_Kaplan}:
\begin{eqnarray}
x^{\mu} \left(X^{\alpha}\right) &=& b^{\mu} + \Lambda^{\mu}_{\alpha}\left(t\right)\,X^{\alpha}\,,
\label{instantaneous_Poincare_transformation}
\end{eqnarray}

\noindent
with $\Lambda^0_0 \left(t\right) = \gamma\left(t\right)$, $\Lambda^i_0 \left(t\right)
= \Lambda_i^0 \left(t\right) =
\gamma\left(t\right)\,\frac{\displaystyle v_i\left(t\right)}{\displaystyle c}$,
$\Lambda_i^j \left(t\right) =
\delta_{ij} + \left(\gamma\left(t\right)-1\right)\frac{\displaystyle
v_i\left(t\right)\,v_j\left(t\right)}{\displaystyle v\left(t\right)^2}$.
Like in (\ref{Poincare_transformation}) we take $b^{\mu} = \left(b^0,\ve{b}\right)$, and $\ve{b}$ points
from the origin of global frame to the origin of the inertial frame at time $T=0$.

We assume the point-charge Q to be located at the origin of the comoving inertial system  
and then the four-potential in this coordinate system is given by
\begin{eqnarray}
A^{\mu} \left(T,\ve{X}\right) &=& \left(\frac{1}{4\,\pi\,\epsilon_0}\,\frac{Q}{R}\,,\,\ve{0}\right),
\label{ED_2}
\end{eqnarray}

\noindent 
where $R = \left|\ve{X}\right|$, and the four-velocity of the
charge in the local system is $u_Q^{\mu}=\left(c,\ve{0}\right)$.

Now we want to determine the four-potential in the global coordinate system.
As is well-known the Li\'enard-Wiechert potentials for a moving point-charge
expressed in terms of retarded time are independent of acceleration.
Accordingly, it has been argued in \cite{Landau_Lifschitz,Rohrlich}
that one might introduce an instantaneous local
rest-system as described above and with the point-charge at its origin at retarded time  
$t_{\rm ret} = t - \left|\ve{x} - \ve{x}_Q\left(t_{\rm ret}\right)\right|/c$,  
and where the four-potential is given by (\ref{ED_2}).
Then, an instananeous
Poincar\'e transformation (\ref{instantaneous_Poincare_transformation}) at $t
= t_{\rm ret}$ yields  
\begin{eqnarray}
A^{\mu} \left(t,\ve{x}\right) &=& \left. \frac{1}{4\,\pi\,\epsilon_0}\, 
\frac{Q\,u_Q^{\mu}\left(t\right)} {\left| u_{\mu}^Q \left(t\right)
\left(x^{\mu} - x^{\mu}_Q\left(t\right) \right) \right|}\right\vert_{t=t_{\rm ret}}\,, 
\label{ED}
\end{eqnarray}

\noindent
where $u_Q^{\mu}\left(t\right) = \gamma\left(t\right) \left(c,\ve{v}_Q\left(t\right)\right)$ 
is the four-velocity of $Q$ in the global system and  
all time-dependent quantities on the right-hand side have to be taken at retarded time $t_{\rm ret}$.
Furthermore, in (\ref{ED}) the local coordinate distance $R$ has been replaced by the
Lorentz-invariant distance, cf. Eq.~(\ref{invariant_distance_1}): 
\begin{eqnarray}
\rho &=& \left. \frac{\left| u^Q_{\mu}\left(t\right)\,\left(x^{\mu} - x^{\mu}_Q
\left(t\right) \right) \right|}{c} \right\vert_{t=t_{\rm ret}} 
= \left. \gamma\left(t\right)\,\left(r_Q\left(t\right) -
\frac{\ve{v}_Q\left(t\right) \cdot \ve{r}_Q\left(t\right)}{c}\right)\right\vert_{t=t_{\rm ret}}\,,  
\label{invariant_distance_3}
\end{eqnarray}

\noindent
where $\ve{r}_Q\left(t\right) = \ve{x} - \ve{x}_Q\left(t\right)$ and 
$r_Q\left(t\right) = \left|\ve{r}_Q\left(t\right)\right|$.   
The solution (\ref{ED}) which has been obtained by an instananeous Poincar\'e transformation 
is nothing else than the well-known Li\'enard-Wiechert potentials in classical electrodynamics.

\subsection{Arbitrarily moving mass-monopoles}\label{Monopole_without_spin}

Now we are going to determine the metric of a pointlike body A moving arbitrarily along a time-like 
trajectory $x_A^{\mu}\left(T\right)$ in the global system with the aid of the same approach as described 
in the previous section.
According to (\ref{Local_Metric}), the metric of a pointlike body without spin
and in its local rest frame $X^{\mu} = \left(c T,\ve{X}\right)$ is given by
\begin{eqnarray}
H^{\alpha \beta}_{\left(M\right)} \left(T,\ve{X}\right) &=& \frac{2\,G\,M}{c^2\,R}\,\delta_{\alpha\beta}\,,
\label{Arbitrarily_Moving_Body_1}
\end{eqnarray}

\noindent
where $M$ is the mass monopole $M_L$ defined by Eq.~(5.33) in \cite{Multipole_Damour_2} for the special case $l=0$.
For the case of an arbitrarily moving pointlike charge we perform an
instantaneous Poincar\'e transformation (\ref{instantaneous_Poincare_transformation}) of the metric 
field (\ref{Arbitrarily_Moving_Body_1}) at the retarded instant of
time defined by Eq.~(\ref{retarded_time_D}), and obtain the global metric
\begin{eqnarray}
h^{\mu \nu}_{\left(M\right)} \left(t,\ve{x}\right) &=& \frac{2\,G\,M}{c^2}\,
\left. \frac{\Lambda^{\mu}_{\alpha} \left(t\right)\,\Lambda^{\nu}_{\beta} \left(t\right)\,
\delta_{\alpha\beta}}{\gamma\left(t\right)\,\left(r\left(t\right) -
\frac{\displaystyle \ve{v}\left(t\right) \cdot \ve{r}\left(t\right)}{\displaystyle c}\right)}
\right\vert_{t=t_{\rm ret}} \,,
\label{Arbitrarily_Moving_Body_2}
\end{eqnarray}

\noindent
where $\ve{r}\left(t\right) = \ve{x} - \ve{x}_A\left(t\right)$ and
$r\left(t\right) = \left|\ve{r}\left(t\right)\right|$ and 
for the distance $R$ we have used the invariant expression (\ref{invariant_distance_3}). 
Now we use the relation
$c^2\,\Lambda^{\mu}_{\alpha}\;\Lambda^{\nu}_{\beta}\,\delta_{\alpha \beta} = 2\,u^{\mu}\,u^{\nu} + c^2\,\eta^{\mu \nu}$,
and obtain
\begin{eqnarray}
h^{\mu \nu}_{\left(M\right)} \left(t,\ve{x}\right) &=& \frac{4\,G\,M}{c^2}\,
\left. \frac{1}{\gamma\left(t\right)\,\left(r\left(t\right) -
\frac{\displaystyle \ve{v}\left(t\right) \cdot \ve{r}\left(t\right)}{\displaystyle c}\right)}\,
\left(\frac{u^{\mu}\left(t\right)}{c}\,\frac{u^{\nu}\left(t\right)}{c}+\frac{\eta^{\mu \nu}}{2}\right)
\right\vert_{t=t_{\rm ret}}\,, 
\label{Arbitrarily_Moving_Body_4}
\end{eqnarray}

\noindent 
where $u^{\mu}\left(t\right) = \gamma\left(t\right)\left(c, \ve{v}\left(t\right)\right)$  
is the four-velocity of the body and 
$\ve{v}\left(t\right)$ being the three-velocity of the body in the global system.  
The expression (\ref{Arbitrarily_Moving_Body_4}) is the contribution to the metric of one arbitrarily moving
pointlike body in post-Minkowskian approximation. The metric for the case of N pointlike bodies 
is simply obtained by a summation over N individual contributions (\ref{Arbitrarily_Moving_Body_4}),   
in agreement with Eq.~(10) in \cite{KopeikinSchaefer1999} or Eq.~(11) in \cite{KopeikinMashhoon2002}.  

For many situations, the slow-motion approximation ($v \ll c$) is of sufficient accuracy, e.g. \cite{Bruegmann_Paper, Deng_Xie}.  
Therefore, we will compare the metric (\ref{Arbitrarily_Moving_Body_4}) with previous results 
in the literature in the slow-motion approximation. A corresponding series expansion of (\ref{Arbitrarily_Moving_Body_4}) yields  
\begin{eqnarray}
h^{00}_{\left(M\right)} \left(t,\ve{x}\right) &=& \frac{2\,G\,M}{c^2}\,\frac{1}{r\left(t\right)}\left.
\left(1 + \frac{\ve{v}\left(t\right)\cdot\ve{r}\left(t\right)}{c\,r\left(t\right)}
+ \frac{\left(\ve{v}\left(t\right)\cdot\ve{r}\left(t\right)\right)^2}
{c^2\,r^2\left(t\right)}
+ \frac{3}{2}\,\frac{v^2\left(t\right)}{c^2}\right)\right\vert_{t=t_{\rm ret}}
+ {\cal O}\left(c^{-5}\right), 
\nonumber\\
\label{Comparison_5}
\\
\nonumber\\
h^{0i}_{\left(M\right)} \left(t,\ve{x}\right) &=& \frac{4\,G\,M}{c^2}\,\frac{1}{r\left(t\right)}  
\frac{v_i\left(t\right)}{c}\,\left.
\left(1 + \frac{\ve{v}\left(t\right)\cdot\ve{r}\left(t\right)}{c\,r\left(t\right)} \right)
\right\vert_{t=t_{\rm ret}} + {\cal O}\left(c^{-5}\right),
\label{Comparison_10}
\\
\nonumber\\
h^{ij}_{\left(M\right)} \left(t,\ve{x}\right) &=& 
\frac{2\,G\,M}{c^2}\,\frac{1}{r\left(t\right)}\,\delta_{ij} \left.
\left(1 + \frac{\ve{v}\left(t\right)\cdot\ve{r}\left(t\right)}{c\,r\left(t\right)}
+ \frac{\left(\ve{v}\left(t\right)\cdot\ve{r}\left(t\right)\right)^2}
{c^2\,r^2\left(t\right)}
- \frac{1}{2}\,\frac{v^2\left(t\right)}{c^2} \right)\right\vert_{t=t_{\rm ret}} 
\nonumber\\ 
\nonumber\\ 
&& + \frac{4\,G\,M}{c^2}\,\frac{1}{r\left(t\right)}\,
\left. \frac{v_i\left(t\right)\,v_j\left(t\right)}{c^2}  
\right\vert_{t=t_{\rm ret}} + {\cal O}\left(c^{-5}\right).
\label{Comparison_15}
\end{eqnarray}

\noindent
The retarded time-argument in (\ref{Comparison_5}) - (\ref{Comparison_15})  
has to be replaced by the global coordinate time using the following relations: 
\begin{eqnarray}
\ve{r} \left(t_{\rm ret}\right) &=& \ve{r}\left(t\right) + \frac{\ve{v}\left(t\right)}{c}\,
r\left(t\right) + \frac{\ve{v}\left(t\right)}{c}\,
\frac{\ve{v}\left(t\right) \cdot \ve{r}\left(t\right)}{c}  
+ {\cal O}\left(c^{-3}\right) + {\cal O}\left(G\right),
\label{t_ret_5}
\\
\nonumber\\
r \left(t_{\rm ret}\right) &=& r\left(t\right)
\left(1 + \frac{\ve{r} \left(t\right) \cdot \ve{v}\left(t\right)}{c\,r\left(t\right)}
+ \frac{1}{2}\,\frac{v^2\left(t\right)}{c^2} + \frac{1}{2}\,
\frac{\left(\ve{v}\left(t\right) \cdot \ve{r}\left(t\right)\right)^2}{c^2\,r^2 \left(t\right)}\right)
+ {\cal O}\left(c^{-3}\right) + {\cal O}\left(G\right), 
\label{t_ret_10}
\\
\nonumber\\
\frac{v_i \left(t_{\rm ret}\right)}{c} &=& \frac{v_i\left(t\right)}{c} + {\cal O}\left(c^{-3}\right) + {\cal O} \left(G\right),
\label{t_ret_15}
\end{eqnarray}

\noindent
where we have taken into account that for a system of N pointlike masses the 
acceleration is proportional to gravitational constant due to the equations of motion; 
see also text below Eq.~(23) in \cite{Bruegmann_Paper}. 
Then, to order $G$ we obtain:  
\begin{eqnarray}
h^{00}_{\left(M\right)} \left(t,\ve{x}\right) &=& \frac{2\,G\,M}{c^2}\,\frac{1}{r\left(t\right)}
\left(1 - \frac{1}{2}\,\frac{\left(\ve{v}\left(t\right)\cdot\ve{r}\left(t\right)\right)^2}
{c^2\,r^2\left(t\right)}
+ 2\,\frac{v^2\left(t\right)}{c^2}\right) + {\cal O}\left(c^{-5}\right),
\label{Comparison_20}
\\
\nonumber\\
h^{0i}_{\left(M\right)} \left(t,\ve{x}\right) &=& \frac{4\,G\,M}{c^2}\,\frac{1}{r\left(t\right)}  
\frac{v_i\left(t\right)}{c} + {\cal O}\left(c^{-5}\right)\,, 
\label{Comparison_25}
\\
\nonumber\\
h^{ij}_{\left(M\right)} \left(t,\ve{x}\right) &=&
\frac{2\,G\,M}{c^2}\,\frac{1}{r\left(t\right)} 
\left(\delta_{ij} 
- \frac{1}{2}\,\frac{\left(\ve{v}\left(t\right)\cdot\ve{r}\left(t\right)\right)^2}
{c^2\,r^2\left(t\right)}\,\delta_{ij}    
+ 2\,\frac{v_i\left(t\right)\,v_j\left(t\right)}{c^2}\right) + {\cal O}\left(c^{-5}\right)\,,
\label{Comparison_30}
\end{eqnarray}

\noindent
which agrees with Eqs.~(21) - (23) in \cite{Bruegmann_Paper} or with   
Eqs.~(47) - (49) in \cite{Deng_Xie} (for $\beta=\gamma=\epsilon=1$ in \cite{Deng_Xie});  
recall $h^{0i} = - h_{0i}$, while $h^{00}=h_{00}$ and $h^{ij}=h_{ij}$, and  
all relations are valid only to first order in $G$.

\subsection{Arbitrarily moving Spin-dipoles}\label{Spin}

Now we proceed with the consideration of the metric of a pointlike body with spin. 
According to (\ref{Local_Metric}), the metric of a massive body with monopole and
spin is, in its local rest frame $X^{\mu} = \left(c T,\ve{X}\right)$, given by
\begin{eqnarray}
H^{\alpha \beta} \left(T,\ve{X}\right) &=&
H^{\alpha \beta}_{\left(M\right)} \left(T,\ve{X}\right) + H^{\alpha \beta}_{\left(S\right)} \left(T,\ve{X}\right),
\label{Arbitrarily_Moving_Body_Spin_1}
\end{eqnarray}

\noindent
where the monopole part has been given by Eq.~(\ref{Arbitrarily_Moving_Body_1}) and the spin part is given by
\begin{eqnarray}
H^{0 a}_{\left(S\right)} \left(T,\ve{X}\right) &=& - \frac{4\,G}{c^3}\,\epsilon_{a\,b\,c}\,
\frac{\partial}{\partial X^b}\,\frac{S_c}{R}\,,
\label{Arbitrarily_Moving_Body_Spin_2}
\end{eqnarray}

\noindent
while all other components of the spin part vanish: $H^{00}_{\left(S\right)} = 0$ and $H^{ij}_{\left(S\right)} = 0$.
Again, for simplicity we assume here that the co-moving system is located at the center of mass 
of this body, and we neglect the time-dependence of the spin vector in the local system.  

Now we perform an instantaneous Poincar\'e transformation of the local metric
(\ref{Arbitrarily_Moving_Body_Spin_2}), and obtain the spin part in global coordinates for an
arbitrarily moving pointlike body with spin:
\begin{eqnarray}
h^{\alpha \beta}_{\left(S\right)} \left(t,\ve{x}\right) &=&
\frac{4\,G}{c^3}\,\Lambda_0^{( \mu}\left(t_{\rm ret}\right)\,\Lambda^{\nu
)}_a \left(t_{\rm ret}\right)\, \epsilon_{a\,b\,c}\,\Lambda^{\lambda}_b
\left(t_{\rm ret}\right)\,S_c\, \frac{\partial}{\partial
x^{\lambda}}\,\frac{1}{\rho}\,, \label{Arbitrarily_Moving_Body_Spin_5}
\end{eqnarray}

\noindent where for $R$ we have used the invariant expression $\rho$ given by
(\ref{invariant_distance_1}) for the distance. By performing the very same
steps as described in some detail in section \ref{spin_uniform}, we obtain 
for (\ref{Arbitrarily_Moving_Body_Spin_5}) the following expression:
\begin{eqnarray}
h^{\mu \nu}_{\left(S\right)} \left(t,\ve{x}\right) &=& \frac{4\,G}{c^4}\;
\left. \frac{r_{\alpha}\left(t\right)\,S^{\alpha ( \mu}_{\rm global}\,u^{\nu )}\left(t\right)}
{\gamma^3\left(t\right)\left(r\left(t\right)
-\frac{\displaystyle \ve{v}\left(t\right)\cdot\ve{r}\left(t\right)}{\displaystyle c}\right)^3}
\right\vert_{t=t_{\rm ret}}\,.
\label{Arbitrarily_Moving_Body_Spin_6}
\end{eqnarray}

\noindent
Eq.~(\ref{Arbitrarily_Moving_Body_Spin_6}) is the result for the spin part of the metric of one arbitrarily moving pointlike
massive body with spin,  
cf. Eq.~(16) in \cite{KopeikinMashhoon2002}. Recall, that (\ref{Spin_mu_nu}) was
valid for the case of an extended body but in uniform motion.
Like in the previous section, the metric of a system of N arbitrarily moving pointlike 
spin-dipoles is simply obtained by a summation 
over the contributions (\ref{Arbitrarily_Moving_Body_Spin_6}) of N individual pointlike spin-dipoles.   
In many situations, the metric for a spinning body in slow-motion ($v \ll c$) is sufficient, 
e.g. \cite{Xie}. Hence, 
like for the case of pointlike monopoles, we will compare (\ref{Arbitrarily_Moving_Body_Spin_6}) 
with results previously obtained in the literature in the slow-motion approximation.  
By inserting (\ref{t_ret_5}) - (\ref{t_ret_15}) into (\ref{Arbitrarily_Moving_Body_Spin_6}) we obtain 
\begin{eqnarray}
h^{00}_{\left(S\right)} \left(t,\ve{x}\right) &=& - \frac{4\,G}{c^4}\,\frac{1}{r^3\left(t\right)}\,r_a\left(t\right)\,S_b\, 
\epsilon_{abc}\,v_c\left(t\right) + {\cal O} \left(c^{-5}\right),  
\label{Comparison_Spin_5}
\\
\nonumber\\
h^{0i}_{\left(S\right)} \left(t,\ve{x}\right) &=& - \frac{2\,G}{c^3}\,\frac{1}{r^3\left(t\right)}\,r_a\left(t\right)\,
\epsilon_{iab}\,S_b\,
+ {\cal O} \left(c^{-5}\right),
\label{Comparison_Spin_10}
\\
\nonumber\\
h^{ij}_{\left(S\right)} \left(t,\ve{x}\right) &=& - \frac{4\,G}{c^4}\,\frac{1}{r^3\left(t\right)}\,r_a\left(t\right)\,S_b\,  
\epsilon_{ab\,(\,i}\,v_{j\,)}\left(t\right) + {\cal O} \left(c^{-5}\right),  
\label{Comparison_Spin_15}
\end{eqnarray}

\noindent
in agreement with Eqs.~(C.17) - (C.19) in \cite{Xie}.
Recall, that (\ref{Comparison_Spin_10}) generates the Lense-Thirring effect, 
the spin in the local frame is time-independent,   
and all relations are valid to first order in $G$.

\section{Conclusions}\label{Conclusions}

Extremely high precision astrometry, high precision space missions and
certain tests of General Relativity, require the knowledge of the metric
tensor of the solar system, or more generally, of a gravitational N-body
system in post-Minkowskian approximation. So far, the metric outside of
massive and moving bodies in only known in post-Newtonian approximation. In
our study, we have considered the metric of massive bodies in motion in
post-Minkowskian approximation, that is valid to any order in velocity
$\displaystyle v/c$. Two different scenarios were on the scope of our
investigation: (i) the case of one body with full mass and spin
multipole structure in uniform motion ($\ve{v}={\rm const}$) in
post-Minkowskian approximation, and (ii) the case of N arbitrarily moving
pointlike bodies with time-dependent speed $\ve{v}\left(t\right)$ in
post-Minkowskian approximation.

For the first problem, a co-moving inertial system of coordinates has been introduced and
the starting point is the local metric given in terms of Damour-Iyer moments.
A Poincar\'e transformation then yields the metric tensor in the
global system (\ref{Global_Metric_A}) in post-Minkowskian approximation.
We have demonstrated that our results are in agreement with known results for pointlike masses having
monopole and spin structure and moving uniformly.

Then we have derived the global metric for pointlike massive bodies in arbitrary motion having monopole  
structure (\ref{Arbitrarily_Moving_Body_4}) and spin structure (\ref{Arbitrarily_Moving_Body_Spin_6}).
We have shown that our results are exact to post-Minkowskian order for the problem of pointlike 
mass-monopoles and spin-dipoles in arbitrary motion.  

The problem to find a global metric for a system of N arbitrarily moving and 
arbitrarily shaped bodies in post-Minkowskian approximation is highly complex  
and one encounters many subtle difficulties.  
Especially (in contrast to the case of pointlike bodies), such a metric cannot be obtained by a simple 
instantaneous Poincar\'e transformation of the metric (\ref{Local_Metric}) for extended bodies.  
Moreover, it is obvious that for this problem
a corresponding accelerated local reference system has to be constructed. It
is clear that such a local system can be defined in many different ways
(e.g., Fermi normal coordinates or special harmonic ones). As is well known,
however, that even in the case of vanishing gravitational fields, i.e., in
Minkowski space, such a construction is highly problematic; the reader is
referred to
\cite{Manasse_Misner,Marsh,Desloge,Ni-Zimm,MarzlinA,MarzlinB,Pauri,Gourgoulhon}.
At the moment being, we consider our study as one more step towards the aim of a global metric for
a system of N arbitrarily shaped and arbitrarily moving massive bodies in
post-Minkowskian approximation.

\section*{Acknowledgment}

The authors acknowledge detailed and fruitful discussions with  
Professor Sergei A. Klioner.
This work was supported by the Deutsche Forschungsgemeinschaft (DFG).

\appendix

\section{Notation for the metric tensors\label{Metric_Notation}}

All relations given here will be valid to first order in $G$, without 
explicit indication.
For weak gravitational fields the metric differs only slightly from flat space metric,  
that means
\begin{eqnarray}
G_{\mu \nu} &=& \eta_{\mu \nu} + H_{\mu\nu}\;, \quad  
G^{\mu \nu}=\eta^{\mu \nu} - H^{\mu\nu}\;,
\label{metric_tensor_5}
\end{eqnarray}

\noindent
where $\eta_{\mu \nu} = \eta^{\mu \nu}$ is the metric of Minkowski space, 
and $\left|H_{\mu\nu}\right| \ll 1$ and $\left|H^{\mu\nu}\right| \ll 1$. 

The equations of linearized gravity take a simple form in the gothic metric 
\cite{Landau_Lifschitz,Kopeikin_Efroimsky_Kaplan,Gothic_Metric}, defined by  
\begin{eqnarray}
\frac{G_{\mu \nu}}{\sqrt{-G}} &=& \eta_{\mu \nu} + \overline{H}_{\mu\nu}\;,  
\quad \sqrt{-G}\,G^{\mu \nu} = \eta^{\mu \nu} - \overline{H}^{\mu\nu}\,, 
\label{metric_tensor_10}
\end{eqnarray}

\noindent  
where $G = {\rm det} \left(G_{\mu \nu}\right)$ is the determinant of metric tensor. 
The factor $\sqrt{-G}$ implies that the gothic metric is not a tensor but a tensor density.
Let us further note the following relations for the trace-reversed metric perturbation:
\begin{eqnarray}
\overline{H}_{\mu \nu} &=& H_{\mu \nu} - \frac{1}{2}\,\eta_{\mu\nu}\,H\;, \quad 
\overline{H}^{\mu \nu} = H^{\mu \nu} - \frac{1}{2}\,\eta^{\mu\nu}\,H\;, 
\label{metric_tensor_15}
\end{eqnarray}

\noindent
where $H = \eta^{\alpha \beta}\,H_{\alpha \beta}$. The inverse relation reads  
\begin{eqnarray}
H_{\mu \nu} &=& \overline{H}_{\mu \nu} - \frac{1}{2}\,\eta_{\mu\nu}\,\overline{H}\;, \quad 
H^{\mu \nu} = \overline{H}^{\mu \nu} - \frac{1}{2}\,\eta^{\mu\nu}\,\overline{H}\;, 
\label{metric_tensor_20}
\end{eqnarray}

\noindent
where $\overline{H} = \eta^{\alpha \beta}\,\overline{H}_{\alpha \beta}$.
Finally we note $H = - \overline{H}$, and we find  
\begin{eqnarray}
\sqrt{-G} &=& 1 - \frac{1}{2}\,\overline{H} \;, \quad \sqrt{-G} = 1 + \frac{1}{2}\,H\;. 
\label{metric_tensor_30}
\end{eqnarray}

\section{Some relations for the Spin}\label{Spin_Tensor_Relations}

\subsection{Lorentz transformation of Spin}

In the local frame the spin four-vector is denoted by ${S}^{\mu} = \left(0,\ve{S}\right)$, while in the global
system the spin four-vector is denoted by $S_{\rm global}^{\mu} = \left(S^0_{\rm global},\ve{S}_{\rm global}\right)$.
The Lorentz transformation for the spin between the co-moving frame co-moving with the massive body and the global frame
reads
\begin{eqnarray}
{S}^i_{\rm global} &=& \Lambda^i_{\mu}\,{S}^{\mu}
= {S}^i + \left(\gamma - 1\right)\,\frac{\ve{v} \cdot \ve{S}}{v^2}\,v_i\,,
\label{Lorentz_Spin_1}
\\
\nonumber\\
S^0_{\rm global} &=& \Lambda^0_{\mu}\,{S}^{\mu} = \gamma \left(\frac{\ve{v} \cdot \ve{S}}{c}\right)\,.
\label{Lorentz_Spin_2}
\end{eqnarray}

\noindent
Note, that the spin four-vector in any Lorentz frame has three independent components only.
The transformation (\ref{Lorentz_Spin_1}) and (\ref{Lorentz_Spin_2}) agree with Eq.~(8) in \cite{KopeikinMashhoon2002}. The
inverse transformation can easily be deduced from Eqs.~(\ref{Lorentz_Spin_1}) and (\ref{Lorentz_Spin_2}) and is given by
\begin{eqnarray}
{S}^i &=& S^i_{\rm global} + \frac{1-\gamma}{v^2}\,\frac{c}{\gamma}\,S^0_{\rm global}\,v_i\,.
\label{Lorentz_Spin_3}
\end{eqnarray}

\noindent
Of course, relation (\ref{Lorentz_Spin_3}) can also be obtained from the inverse Lorentz transformation.

\subsection{Proof of relation (\ref{Spin_00_10})}

In \cite{KopeikinMashhoon2002}, some anti-symmetric spin tensor in global coordinates $S^{\alpha \beta}_{\rm global}$
has been employed. Due to the anti-symmetry of this tensor and because of the orthogonality relation
$S^{\alpha \beta}_{\rm global}\,u_{\beta}=0$, this spin tensor has three independent degrees of freedom like the
spin four-vector $S^{\mu}_{\rm global}$, thus both mathematical expressions are on an equal footing. Therefore, the
anti-symmetric spin tensor $S_{\rm global}^{\alpha\,\beta}$ and the spin four-vector $S_{\mu}^{\rm global}$ in global
coordinates are related to each other by the following relation, cf. Eq.~(5) in \cite{KopeikinMashhoon2002}
and cf. Eq.~(3.9) in \cite{Bruegmann_Thesis}:
\begin{eqnarray}
{S}^{\alpha\,\beta}_{\rm global} &=& \eta^{\alpha\,\beta\,\gamma\,\delta}\,S_{\delta}^{\rm global}\;
\frac{u_{\gamma}}{c}\,,
\label{appendix_spin_relations_5}
\\
\nonumber\\
{S}_{\alpha}^{\rm global} &=& \frac{1}{2}\,\eta_{\alpha\,\beta\,\gamma\,\delta}\,\frac{u^\beta}{c}\,
S^{\gamma\,\delta}_{\rm global}\,,
\label{appendix_spin_relations_6}
\end{eqnarray}

\noindent
where (\ref{appendix_spin_relations_6}) is the inverse of (\ref{appendix_spin_relations_5}).
Here, $\displaystyle \eta^{\alpha \beta \gamma \delta}= - \frac{1}{\sqrt{-g}}\,\epsilon_{\alpha \beta \gamma \delta}$ and
$\eta_{\alpha \beta \gamma \delta} = \sqrt{-g}\,\epsilon_{\alpha \beta \gamma \delta}$ are the contravariant and
covariant components of the  
Levi-Civita tensor, respectively, and $\epsilon_{\alpha \beta \gamma \delta}$ is the Minkowskian Levi-Civita tensor
with $\epsilon_{0 1 2 3}=1$. Let us note the following relations of this tensor:
\begin{eqnarray}
\epsilon^{i\,j\,0\,k} &=& \epsilon^{0\,i\,j\,k} = - \epsilon_{0\,i\,j\,k} = - \epsilon_{i\,j\,k}\,.
\label{appendix_spin_relations_16}
\end{eqnarray}

\noindent
In harmonic coordinates $g = - 1 + {\cal O}\left(G\right)$, we obtain from Eq.~(\ref{appendix_spin_relations_5}),
up to order ${\cal O} \left(G\right)$,
\begin{eqnarray}
S^{a\,0}_{\rm global} &=& \epsilon^{a\,0\,\gamma\,\delta}\;\frac{u_{\gamma}}{c}\;S_{\delta}^{\rm global}
= \epsilon^{a\,0\,k\,l}\;\frac{u_{k}}{c}\;S_{l}^{\rm global}\,.
\label{appendix_spin_relations_20}
\end{eqnarray}

\noindent
And by means of (\ref{appendix_spin_relations_16}) we finally arrive at
\begin{eqnarray}
S^{a\,0}_{\rm global} &=& \epsilon_{a\,k\,l}\;\gamma\;\frac{v_{k}}{c}\;S_{l}^{\rm global}
= \gamma\;\left(\frac{\ve{v} \times \ve{{S}}}{c}\right)^a\,,
\label{appendix_spin_relations_25}
\end{eqnarray}

\noindent
where in the last term we have used (\ref{Lorentz_Spin_1}), i.e.
$\ve{v} \times \ve{S}_{\rm global} = \ve{v} \times \ve{S}$. Eq.~(\ref{appendix_spin_relations_25}) 
is nothing but
relation (\ref{Spin_00_10}); cf. Eq.~(D1) in \cite{KopeikinMashhoon2002}.

\section{Prof of relation (\ref{Spin_a0_25})}\label{Proof_Relation_Spin_a0_25}

In order to show (\ref{Spin_a0_25}), we insert into Eq.~(\ref{Spin_a0_15}) the explicit form for the Lorentz matrix
and obtain
\begin{eqnarray}
h^{a 0}_B \left(t,\ve{x}\right) &=& - \frac{2\,G}{c^3}\,\gamma
\left(\epsilon_{ija}\,S_j + \left(\gamma - 1\right)\frac{v_a\,v_k}{v^2}\,\epsilon_{ijk}\,S_j\right)\,\frac{X_i}{\rho^3}\,,
\label{appendix_Spin_a0_25_5}
\end{eqnarray}

\noindent
where we have used the abbreviation
\begin{eqnarray}
X_i &=& r_i \left(t_{\rm ret}\right)+\left(\gamma-1\right)\,
\frac{v_i}{v^2}\,\left(\ve{v}\cdot\ve{r}\left(t_{\rm ret}\right)\right)
- \gamma\,r\left(t_{\rm ret}\right)\frac{v_i}{c}\,.
\label{appendix_Spin_a0_25_10}
\end{eqnarray}

\noindent
The metric (\ref{appendix_Spin_a0_25_5}) is still given in terms of the local spin $S^{\mu} = \left(0,\ve{S}\right)$
comoving with the massive body, and we have to transform it into the spin tensor in global coordinates. For the first term
in the parentheses of Eq.~(\ref{appendix_Spin_a0_25_5}) we will use the following relation 
(a proof is given below):  
\begin{eqnarray}
\gamma\,\epsilon_{ija}\,S_j &=& S_{\rm global}^{a i} + \frac{\gamma - 1}{v^2}\,\left(\ve{v}\cdot \ve{S}\right)
\epsilon_{aij}\,v_j\,,
\label{appendix_Spin_a0_25_15}
\end{eqnarray}

\noindent
while for the second term in the parentheses of Eq.~(\ref{appendix_Spin_a0_25_5}) we will use relation (\ref{Spin_00_10}),
and then we obtain
\begin{eqnarray}
h^{a 0}_{B} \left(t,\ve{x}\right) &=& - \frac{2\,G}{c^3}\,
\left(S^{a\,i}_{\rm global} + \frac{\gamma - 1}{v^2}\;\frac{c}{\gamma}\;S_{\rm global}^0\;\epsilon_{a\,i\,j}\;v_j
+ \frac{1 - \gamma}{v^2}\;v_a\;c\;S^{i\,0}_{\rm global}\right)\,\frac{X_i}{\rho^3}\,.
\label{appendix_Spin_a0_25_20}
\end{eqnarray}

\noindent
For the last term in (\ref{appendix_Spin_a0_25_15}) we have also used relation (\ref{Lorentz_Spin_2}).
The metric (\ref{appendix_Spin_a0_25_20}) is now given in terms of global spin variables. But we still have to express
the second term in (\ref{appendix_Spin_a0_25_20}) by the global spin tensor. Therefore, we use the following relation,
cf. Eq.~(\ref{appendix_spin_relations_6}),
\begin{eqnarray}
S^0_{\rm global} &=& \frac{1}{2}\,\epsilon_{k\,l\,m}\,\frac{u_k}{c}\,S^{l\,m}_{\rm global}\,.
\label{appendix_Spin_a0_25_25}
\end{eqnarray}

\noindent
Inserting (\ref{appendix_Spin_a0_25_25}) into (\ref{appendix_Spin_a0_25_20}) yields (recall the anti-symmetry of spin-tensor):
\begin{eqnarray}
h^{a 0}_B \left(t,\ve{x}\right) &=& - \frac{2\,G}{c^3}\,\frac{X_i}{\rho^3}
\nonumber\\
\nonumber\\
&& \hspace{-2.0cm} \times \bigg(S^{a\,i}_{\rm global} + \frac{\gamma - 1}{v^2}\;v_a\;v_b\;S^{i\,b}_{\rm global}
+ \left(\gamma - 1\right) S^{a\,i}_{\rm global} + \frac{\gamma - 1}{v^2}\;v_i\;v_b\;S^{b\,a}_{\rm global}
+ \frac{1 - \gamma}{v^2}\;v_a\;c\;S^{i\,0}_{\rm global}\bigg),
\nonumber\\
\label{appendix_Spin_a0_25_30}
\end{eqnarray}

\noindent
where for the second term in the parentheses of Eq.~(\ref{appendix_Spin_a0_25_20}) after inserting
(\ref{appendix_Spin_a0_25_25}) we have used
\begin{eqnarray}
\epsilon_{a i j}\,\epsilon_{k l m} =
\left| \begin{array}[c]{l}
\delta_{ak} \;\;\;\delta_{al}\;\;\;\delta_{am}
\nonumber\\
\nonumber\\
\delta_{ik}\;\;\;\delta_{il}\;\;\;\delta_{im}
\nonumber\\
\nonumber\\
\delta_{jk}\;\;\;\delta_{jl}\;\;\;\delta_{jm}
\end{array}
\right|\,.
\label{appendix_Spin_a0_25_35}
\end{eqnarray}

\noindent
We recognize that the second and last term in the parentheses of Eq.~(\ref{appendix_Spin_a0_25_30}) cancel each other, as
one can see by using the relation $v_b\,S^{a\,b}_{\rm global} = c\;S^{a\,0}_{\rm global}$ due to
$S^{\alpha\,\beta}_{\rm global}\,u_{\beta} = 0$. For the fourth term in the parentheses of
Eq.~(\ref{appendix_Spin_a0_25_30}) we use $v_b\,S^{b\,a}_{\rm global} = - c\,S^{a\,0}_{\rm global}$ and obtain
\begin{eqnarray}
h^{a 0}_{B} \left(t,\ve{x}\right) &=& - \frac{2\,G}{c^3}\,\frac{X_i}{\rho^3}\;\left(\gamma\;S^{a\,i}_{\rm global}
+ \frac{1 - \gamma}{v^2}\,v^i\;c\;S^{a\,0}_{\rm global}\right).
\label{appendix_Spin_a0_25_40}
\end{eqnarray}

\noindent
Now we reinsert (\ref{appendix_Spin_a0_25_10}) and obtain, recall $c\,\gamma=u^0$ and $v_i\,S^{a\,i} = c\,S^{a\,0}$,
\begin{eqnarray}
h^{a 0}_{B} \left(t,\ve{x}\right) &=& \frac{2\,G}{c^4}\;
\frac{r_{\gamma}\left(t_{\rm ret}\right)\,S^{\gamma a}_{\rm global}\,u^{0}}{\rho^3}\,,
\label{appendix_Spin_a0_25_45}
\end{eqnarray}

\noindent
where we have used the anti-symmetry of the spin-tensor; note
$r_{\gamma} \left( t_{\rm ret}\right) = \left(- r(t_{\rm ret}), \ve{r} (t_{\rm ret})\right)$,
and $r^{\gamma} = \left(r(t_{\rm ret}), \ve{r} (t_{\rm ret})\right)$. Eq.~(\ref{appendix_Spin_a0_25_45})
is just relation (\ref{Spin_a0_25}).

Finally let us proof relation (\ref{appendix_Spin_a0_25_15}). 
We insert the Lorentz transformations (\ref{Lorentz_Spin_1}) and (\ref{Lorentz_Spin_2})
into relation (\ref{appendix_spin_relations_5}) and obtain up to order $G$,
\begin{eqnarray}
{S}^{i\,j}_{\rm global} &=& \epsilon^{i\,j\,\gamma\,\delta}\;S_{\delta}^{\rm global}\;u_{\gamma}
\nonumber\\
\nonumber\\
&=& \epsilon^{i\,j\,k}\;{S}_k\;\gamma + \epsilon^{i\,j\,k}\;\frac{\gamma - 1}{v^2}
\left(\ve{v} \cdot \ve{S}\right)\;v^k\;\gamma
- \epsilon^{i\,j\,k}\;\left(\frac{\ve{v} \cdot \ve{S}}{c}\right)\;\gamma^2\;\frac{v_k}{c}\;,
\label{Appendix_B_15}
\end{eqnarray}

\noindent
where we have also used $u_0 = - c\,\gamma$, $u^0 = c\,\gamma$ and $u_k = u^k = \gamma\,v_k$;
note $S_0^{\rm global} = - S^0_{\rm global}$, $\epsilon_{i\,j\,k} = \epsilon^{i\,j\,k}$,
and $\epsilon^{0\,i\,j\,k} = - \epsilon_{0\,i\,j\,k}$.
Then, by using the relation
$\left(\gamma - 1\right)\,\gamma - \frac{\displaystyle v^2}{\displaystyle c^2}\,\gamma^2 = 1 - \gamma$,
we obtain from Eq.~(\ref{Appendix_B_15}),
\begin{eqnarray}
{S}^{i\,j}_{\rm global} &=& \gamma\,\epsilon_{i\,j\,k}\;{S}_k
+ \frac{1 - \gamma}{v^2}\,\left(\ve{v} \cdot \ve{S}\right)\,\epsilon_{i\,j\,k}\,v_k\,,
\label{Appendix_B_30}
\end{eqnarray}

\noindent
which is just relation (\ref{appendix_Spin_a0_25_15}); cf. Eq.~(D2) in \cite{KopeikinMashhoon2002}.

\end{document}